\documentclass[twocolumn,superscriptaddress,showpacs,aps,PRL,amsmath,amssymb]{revtex4-1} 
\usepackage{graphicx} 
\usepackage{dcolumn}
\usepackage{bm}
\usepackage{epsfig}
\usepackage[usenames,dvipsnames]{xcolor}

\begin{document}

\title{Intact quasiparticles at an Unconventional Quantum Critical Point} 

  \author{M. L. Sutherland}
\affiliation{Cavendish Laboratory, University of Cambridge, JJ Thomson Ave,
  Cambridge, UK, CB3 0HE}
\author{E. C. T. O'Farrell}
\affiliation{Institute for Solid State Physics, University of Tokyo, Kashiwa,
  Japan 277-8581}
  \author{W.H. Toews}
\affiliation{GWPI and Department of Physics and Astronomy, University of Waterloo, Waterloo, Ontario, Canada, N2L 3G1}
\author{J. Dunn}
\affiliation{GWPI and Department of Physics and Astronomy, University of Waterloo, Waterloo, Ontario, Canada, N2L 3G1}
\author{K. Kuga}
\affiliation{Institute for Solid State Physics, University of Tokyo, Kashiwa,
  Japan 277-8581}
\author{S. Nakatsuji}
\affiliation{Institute for Solid State Physics, University of Tokyo, Kashiwa,
  Japan 277-8581} 
  \author{Y. Machida}
\affiliation{Department of Physics, Tokyo Institute of Technology, Meguro, Japan 152-8551} 
  \author{K. Izawa}
\affiliation{Department of Physics, Tokyo Institute of Technology, Meguro, Japan 152-8551} 
\author{R.W. Hill}
\affiliation{GWPI and Department of Physics and Astronomy, University of Waterloo, Waterloo, Ontario, Canada, N2L 3G1}

\date{\today}
\begin{abstract}
We report measurements of in-plane electrical and thermal transport properties in the limit $T \rightarrow 0$ near the unconventional quantum critical point in the heavy-fermion metal $\beta$-YbAlB$_4$. The high Kondo temperature $T_K$ $\simeq$ 200 K in this material allows us to probe transport extremely close to the critical point, at unusually small values of $T/T_K < 5 \times 10^{-4}$. Here we find that the Wiedemann-Franz law is obeyed at the lowest temperatures, implying that the Landau quasiparticles remain intact in the critical region. At finite temperatures we observe a non-Fermi liquid T-linear dependence of inelastic scattering processes to energies lower than those previously accessed. These processes have a weaker temperature dependence than in comparable heavy fermion quantum critical systems, and suggest a new temperature scale of $T \sim 0.3 K$ which signals a sudden change in character of the inelastic scattering. 

 \end{abstract}

\maketitle

The effect of quantum fluctuations on the properties of matter has become an important area of research motivated by the potential for technological advances through harnessing and manipulating quantum mechanical properties.  A quantum critical point (QCP) arises when a continuous transition between competing order occurs at zero temperature. Here, strong quantum fluctuations can drive the formation of new phases of matter \cite{Mathur01}, and may lead to the breakdown of normal Fermi-liquid behaviour in metals. Much of our understanding of QCPs comes from the study of heavy-fermion compounds, which are canonical systems for investigating antiferromagnetic quantum criticality. An important open question in these materials is whether multiple types of QCPs exist, differentiated by their microscopic behaviour in the critical regime.

The standard model of quantum criticality in metals is the Hertz-Moriya-Millis framework \cite{Moriya73a,*Hertz76,*Millis93}, where the suppression of itinerant antiferromagnetic order results in a paramagnetic state with heavy quasiparticles formed as a result of the Kondo screening of $f$-electron moments by electrons in the conduction band. While this works very well in describing many materials, the presence of localized moments, Fermi surface reconstruction, and diverging effective masses at the magnetic transition in YbRh$_2$Si$_2$ has been interpreted as evidence for a more exotic type of quantum criticality \cite{Gegenwart08}. Alternate frameworks have been proposed where a new energy scale, the `effective Kondo temperature', collapses at the QCP leading to the breakdown of the heavy electron metal \cite{Si13}.  At the so-called Kondo-breakdown QCP, the Kondo effect is destroyed and the entire Fermi surface is destabilized.

Distinguishing between these pictures is a challenging experimental task. One promising approach is to search for the breakdown of the Landau quasiparticle picture via thermal transport measurements, which would provide compelling evidence for the existence of less conventional classes of QCPs. In the Kondo-breakdown model for instance, a fractionalised Fermi liquid \cite{Senthil03} emerges and the presence of additional entropy carriers (spinons) enhances the thermal conductivity above that expected in the normal heavy-fermion metallic state \cite{Kim09}. This should lead to a $T=0$ violation of the Wiedemann-Franz (WF) law which relates the electrical conductivity of a material ($\sigma$) with its thermal conductivity ($\kappa$) via the Sommerfeld value of the Lorenz number ($L_0$):
\begin{equation}
\frac{\kappa}{\sigma T}=\frac{\pi^2 k_B^2}{3e^2} = L_0 = 2.45 \times 10^{-8}\text{~W}\Omega\text{K}^{-2}
\label{eq:WF}
\end{equation}
\noindent In contrast, the quasiparticles near a conventional QCP are expected to remain intact, and the WF law obeyed.

In this letter we report a test of the WF law in the heavy-fermion material $\beta$-YbAlB$_4$, which has the unusual property of lying almost exactly at a QCP at ambient magnetic field and pressure \cite{Nakatsuji08}. Magnetic susceptibility measurements in this material have revealed unusual $T/B$ scaling and an effective mass which diverges as $B^{-1/2}$  \cite{Matsumoto11}. Together with the observation of strong valence fluctuations \cite{Okawa10}, these properties have led to the suggestion that the QCP in $\beta$-YbAlB$_4$ is unconventional in nature. Our thermal transport measurements directly address the question of whether the quasiparticles remain intact in the critical region, and act as an important test for microscopic theories of unconventional quantum criticality in this and other heavy-fermion compounds.

\begin{figure} \centering
               \resizebox{\columnwidth}{!}{      
             \includegraphics{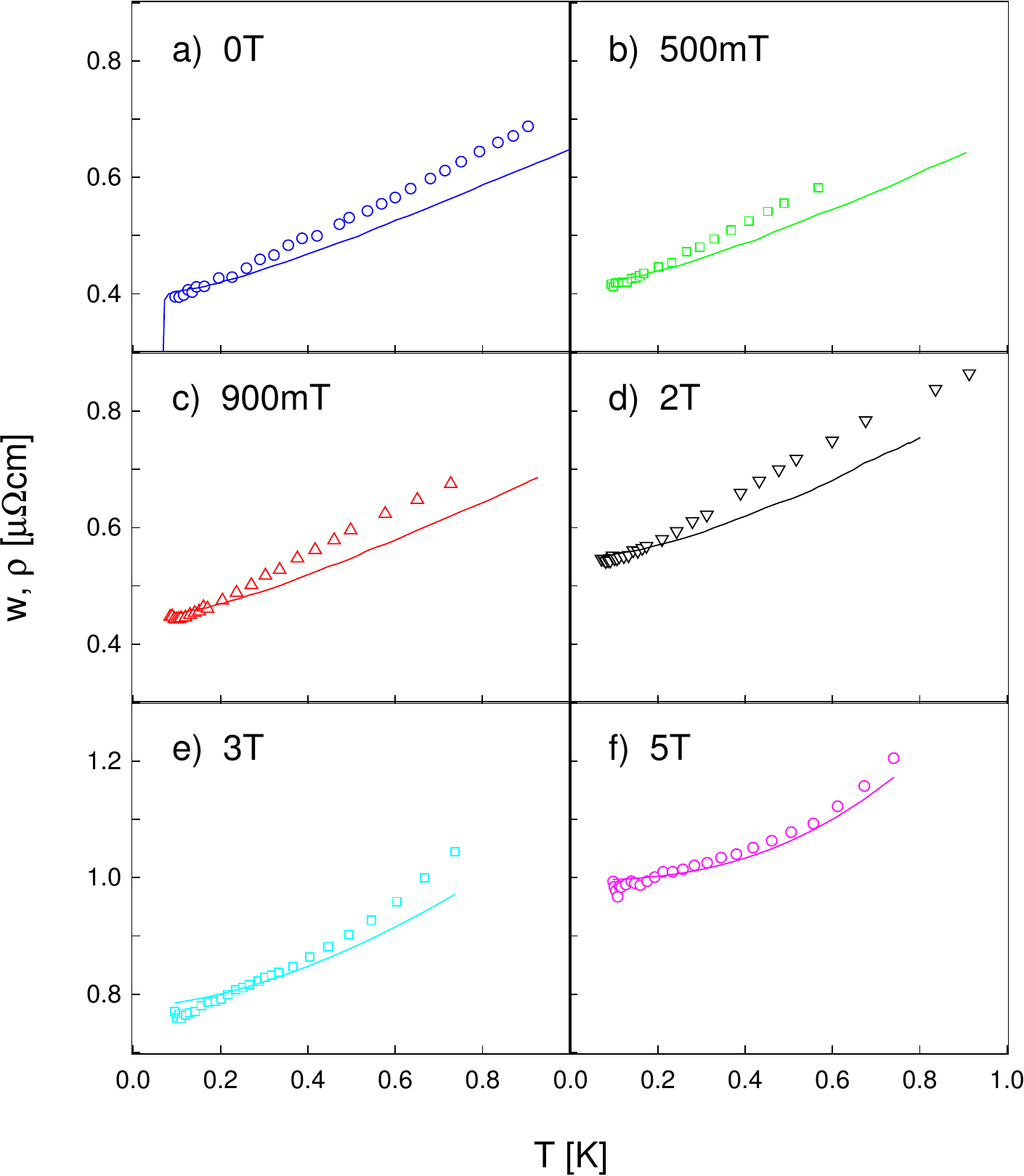}}       
\resizebox{\columnwidth}{!}{
		\includegraphics{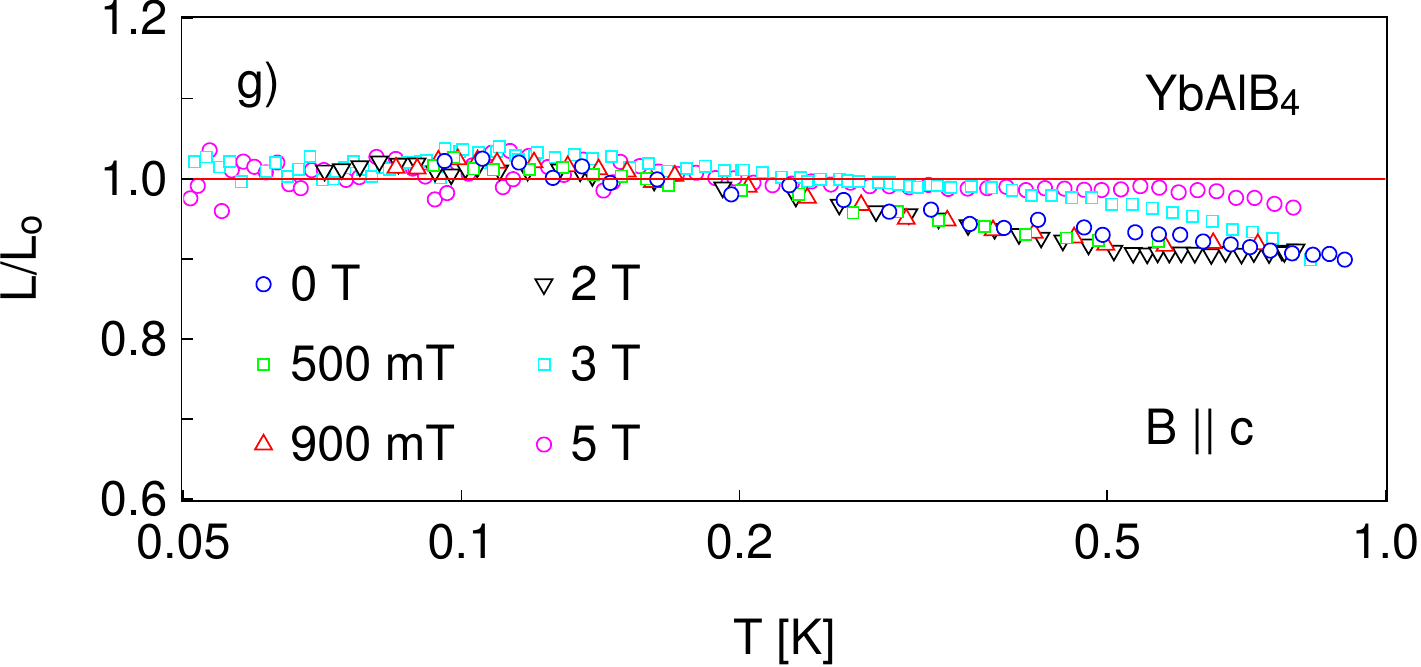}}
              \caption{\label{fig:1} a)-f) In-plane electrical resistivity (lines) and thermal resistivity (open symbols) at various values of applied magnetic field.  g) shows the temperature dependence of the Lorenz ratio $L/L_0$ at each field measured.}            
      \end{figure}

Figure \ref{fig:1} shows the result of thermal and charge transport measurements at several magnetic fields, with $B \parallel c$ and $I \parallel ab$. The high quality single crystals used in this work were thin platelets prepared using a flux growth method \cite{Macaluso07}, with typical sample dimensions of 2mm $\times$ 200$\mu$m $\times$ 15 $\mu$m, and the best sample having a residual resistivity ratio (RRR) of $\rho_{(300~K)}/\rho_{0}$ = 300. Contacts with very low electrical resistance ($<$~5~m$\Omega$) were prepared by firstly ion milling the surface and then depositing Pt contacts. 

The thermal conductivity was measured to $T  \sim$ 70 mK using a two thermometer, one heater steady-state technique, with in-situ thermometer calibrations performed at each field to eliminate the effects of thermometer magnetoresistance. Electrical and thermal transport measurements were taken on the same sample, using the same contacts. In order to compare the charge and heat transport more closely, the thermal conductivity data is converted to thermal resistivity ($w$), in electrical units ($\mu\Omega~\text{cm}$), using Eq.~\ref{eq:WF}: $w = L_0T/\kappa$. We estimate a relative error between thermal and electrical measurements of 3\%, arising mainly from slightly different effective contact separations for heat and charge measurements. Our measurements were repeated in two laboratories using several samples, and we obtain good reproducibility of our results.

In zero field, $\beta$-YbAlB$_4$ lies in a non-Fermi liquid state at low temperatures characterized by an electrical resistivity $\rho(T) \sim T^{1.5}$, a magnetic susceptibility $\chi_c \sim T^{-1/2}$ and an electronic specific heat $\gamma \sim ln(T_0/T)$ with $T_0$ = 200~K, associated with a high-energy valence fluctuation scale \cite{Matsumoto11}. In the cleanest samples, superconductivity is also seen \cite{Nakatsuji08}.

A fit of $B$ = 0 resistivity data in Fig.~\ref{fig:1}a to $\rho(T)=\rho_0+aT^n$ gives $n \sim$ 1.5 (which is identical to that reported previously \cite{Nakatsuji08}), and $\rho_0$ $\sim$ 0.4~$\mu\Omega$cm, indicating the high quality of the samples. We observe a clear drop in resistivity at an onset temperature $T_c$ = 80~mK,  consistent with the presence of superconductivity. A small magnetic field of $B$ = 50~mT completely suppresses superconductivity, consistent with the reported critical field of $B_{c2}$ = 25~mT \cite{Kuga08}. Increasing the field further increases $\rho_0$, (Fig.~\ref{fig:1}b-f), and the temperature dependence increases its curvature, with $\rho(T) \sim T^2$ by $B \geq 3$~T, indicating a return to a Fermi liquid state. Our observations are consistent with the phase diagram for this material elucidated through earlier transport \cite{Nakatsuji08} and magnetization measurements \cite{Matsumoto11}. 

The thermal conductivity at these low temperatures is taken to be dominated by the electron contribution. In a paramagnetic metal, heat is carried by both electrons ($\kappa_{el}$) and phonons ($\kappa_{ph}$), however in the limit $T\rightarrow$~0 samples with low residual resistivities ($\rho_0 < 1~\mu\Omega$cm) should have $\kappa_{el} \gg \kappa_{ph}$, allowing us to ignore the contributions of $\kappa_{ph}$ below 1~K. This is the case in metallic compounds with comparable values of $\rho_0$, for instance YbRh$_2$Si$_2$ \cite{Pfau12}, CeCoIn$_5$ \cite{Tanatar07}, CeRhIn$_5$ \cite{Paglione05} and ZrZn$_2$ \cite{Smith08}.

       \begin{figure*} \begin{center}
\resizebox{18cm}{!}{\includegraphics{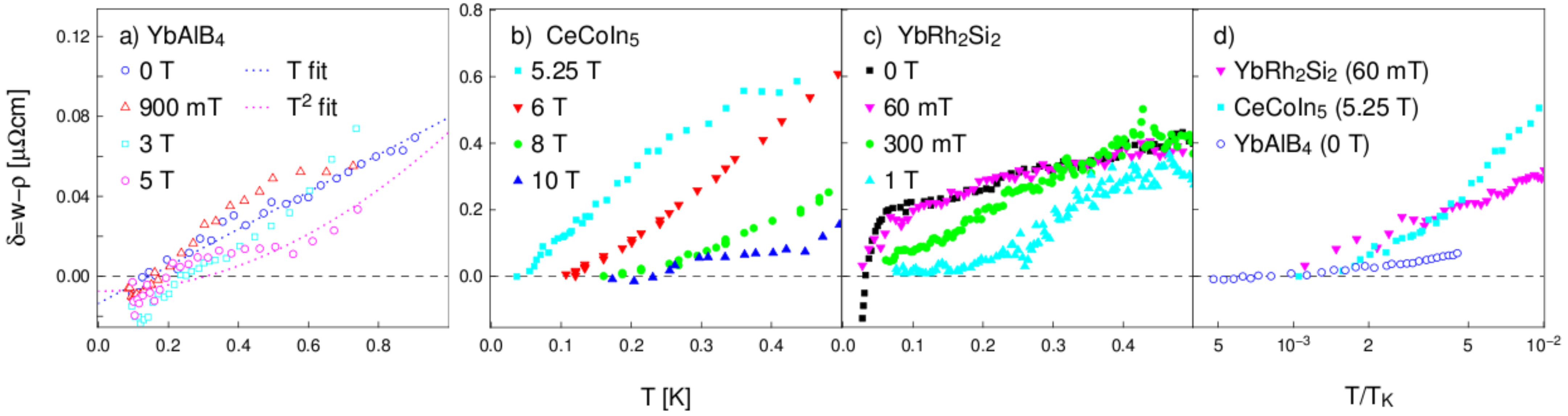}}
              \caption{\label{fig:2} The difference $\delta(T)$ between thermal ($w$) and electrical ($\rho$) in-plane resistivity measured in quantum critical systems a) $\beta$-YbAlB$_4$ b) CeCoIn$_5$ \cite{Tanatar07} and c) YbRh$_2$Si$_2$ \cite{Pfau12}, with fields applied along the $c$-axis. The scale has been reduced in panel a) to make the date for $\beta$-YbAlB$_4$ more clear. The critical fields for these systems are $H_c$ = 0,  5.25~T and 60~mT respectively. The dotted lines in a) show the results of a linear-$T$ fit to the $B$ = 0 data in the non-Fermi liquid state, and a $T^2$ fit to the $B$ = 5T data in the Fermi liquid state. Panel d) shows the data for the three materials with the temperature axis scaled by the respective Kondo temperatures $T_K$ and plotted logarithmically.}             
\end{center}      
\end{figure*}

The comparison of the $w$ and $\rho$ curves in Fig.~\ref{fig:1} offers a direct test of the WF law. 
Clearly, all the $w$ and $\rho$ curves converge below $T$ = 200~mK in both zero and applied magnetic fields \cite{note1}.  From this we conclude that the WF law is robustly obeyed both at the QCP and at all other fields in the phase diagram of $\beta$-YbAlB$_4$, at least to within our experimental resolution of 3\%.  A plot of the Lorenz ratio, $L/L_0$ = $\kappa/\sigma T L_0$, seen in Fig.~\ref{fig:1}g further emphasizes this point, approaching unity as the temperature falls below 200 mK. 

This observation argues against the appearance of a fractionalized Fermi liquid that emerges in some Kondo-breakdown models. Specifically, recent theoretical work predicts that a spinon contribution leads to an \emph{excess} thermal conductivity and a Lorenz ratio that diverges as $L(T) \sim (k_BT)^{-3}$ at low temperatures, before being cut off by elastic scattering as $T\rightarrow$~0 \cite{Hackl11b}. This is expected to give a characteristic peak with $L/L_0 > 1$, which is not observed in $\beta-YbAlB_{4}$.

Our results also stand in stark contrast to recent observations in similar materials. At the putative Kondo-breakdown QCP in YbRh$_2$Si$_2$ for instance, a WF law violation of 10\% has been reported \cite{Pfau12}, with $L/L_0(T \rightarrow 0)\sim$ 0.9 indicating lower than expected heat conduction. However, more recent measurements indicate the WF law is satisfied \cite{Machida:2013fm, Reid2014}.  At the field-tuned magnetic QCP in CeCoIn$_5$, the situation is more complex, showing agreement with the WF law as  $T \rightarrow$~0 for heat current in one direction but violation in another \cite{Tanatar07}.

In this context, our observations are significant. Despite the signatures of unconventional criticality, we find that the quasiparticles which carry heat and charge in $\beta$-YbAlB$_4$ \emph{remain intact at, or near, the unconventional QCP}. It directly follows that  a theoretical treatment of criticality in this material must be able to account for unconventional magnetic and transport properties while maintaining the basic framework of the quasiparticle picture. One recent model that is consistent with our results is proposed by Ramires \textit{et al}., in which a hybridisation gap between conduction and $f$-electons is strongly $k$-dependent, and vanishes along a line in which zero-energy excitations form \cite{Ramires12}.

Having discussed the $T = 0$ results, we now consider the temperature evolution of heat and charge transport.  Deviations of the ratio $L/L_0$ from unity give a measure of the ratio of inelastic to elastic scattering. In Fig.~\ref{fig:1}(g), our data shows that at low temperatures, the ratio of inelastic to elastic scattering processes remains fairly small in $\beta$-YbAlB$_4$, even at the zero field QCP and in very clean samples.  Furthermore, this modest temperature dependence is largely magnetic field independent.  We now analyse these basic observations in more detail.

As $T$ is increased, the thermal and electrical resistivities evolve differently, reflecting dissimilarities in how scattering mechanisms affect heat and charge currents. Consider an electron of mass $m^{\star}$, wavevector $k_F$ and an energy $E$ relative to the Fermi energy. When scattered through an angle $\theta$, the electric current $j_{\rho}$ and thermal current $j_{w}$ are reduced by amounts \cite{Schriempf69}:
\begin{equation}
\Delta j_{\rho} =-\frac{ek_F}{m^{\star}}(1-\cos(\theta))
\label{eq:1}
\end{equation}
\begin{equation}
\Delta j_{w} =-\frac{k_F}{m^{\star}}[E(1-\cos\theta)+\Delta E \cos\theta]
\label{eq:2}
\end{equation}

The $(1-\cos(\theta))$ terms in these equations arise from changes in the direction of the wavevector of the electron, and are often referred to as `horizontal processes', $K_{hor}$. The second term in Eq.~\ref{eq:2} arises due to the fact that the electron may suffer a loss of energy $\Delta E$ in an inelastic collision, which further degrades a heat current. This is often referred to as a `vertical process', $K_{ver}$. In the case of elastic scattering $\Delta E$ = 0 and the WF law is recovered. 

In magnetic scattering, a full calculation of the electrical and thermal resistivities involves integrating Eqs.~\ref{eq:1} and \ref{eq:2} over the $\mathbf{q}$ and $\omega$ dependence of the fluctuation spectrum. Horizontal processes are weighted by a factor of $\mathbf{q}^2$, while vertical processes are weighted by a factor of $\omega^2$ \cite{Schriempf69, Ziman}. Comparing heat and charge resitivities directly can thus give access to detailed information about the $\mathbf{q}$ and $\omega$ dependence of scattering.

Defining $\delta(T)$ = $w(T) - \rho(T)$ yields the temperature dependence of the vertical scattering processes, since $\rho \sim K_{hor}$ and $w \sim K_{hor} + K_{ver}$. In Fig.~\ref{fig:2}, we plot $\delta(T)$ for $\beta$-YbAlB$_4$ alongside equivalent data for the unconventional quantum critical systems CeCoIn$_5$ \cite{Tanatar07} and YbRh$_2$Si$_2$ \cite{Pfau12}. In zero field, a linear fit to $\delta(T)$ in $\beta$-YbAlB$_4$  for $T<0.8$ K is shown to describe the data remarkably well,. This $T-$linear property of inelastic scattering is strikingly different than what is expected in a conventional Fermi-liquid where $\delta(T) \sim T^2$ or $\delta(T) \sim T^3 + T^5$ from electron-electron and electron-phonon scattering respectively \cite{Ziman}.  

A similar linear variation of $\delta(T)$ was previously observed arising from critical ferromagnetic fluctuations in the itinerant $d$-metal ZrZn$_2$ \cite{Sutherland12b}. The high Wilson ratio $R_W$ $\sim$ 7 in $\beta$-YbAlB$_4$ suggests the presence of ferromagnetic fluctuations in the Yb moments \cite{Julian99}.  However, despite evidence that local moments may persist to low temperatures \cite{Matsumoto11}, it is unclear how strongly these local fluctuations might couple to the itinerant electrons. Valence fluctuations are another important consideration. X-ray photoemission spectroscopy measurements reveal a mixed valence of Yb$^{+2.75}$ in $\beta$-YbAlB$_4$ \cite{Okawa10}, associated with Yb$^{+3}$~$\rightleftharpoons$~Yb$^{+2}$ fluctuations that may play a role in the quantum critical behaviour. Such fluctuating modes have been shown to lead to $T$-linear electrical resistivity at intermediate temperatures, and $T^{3/2}$ behaviour at low temperatures \cite{Watanabe12,Miyake14}, similar to that observed in $\beta$-YbAlB$_4$. Whether such valence fluctuations could also give rise to a linear temperature dependence in $\delta(T)$ is an open theoretical question.  

The magnetic field dependence of $\delta(T)$ is also interesting.  For the most part, the approximately linear-in-temperature behaviour is maintained at low field.  As the field increases towards 5 T, the curvature increases towards a $T^2$ dependence, shown by the fit in the figure, and the magnitude at low temperatures is smaller.  This is roughly consistent with the electronic specific heat coefficient $\gamma$ which is reduced by applying a magnetic field at low temperatures \cite{Matsumoto11}.  
It also consistent with $\delta(T)$ in other field-driven quantum critical systems. In Fig.~\ref{fig:2}b for example, applying a magnetic field greater than $H_c$ to CeCoIn$_5$ quenches inelastic scattering, changing $\delta(T)$ from $T$-linear at the critical field to $\delta(T) \sim T^2$ by 10~T, where the Fermi liquid state is recovered  \cite{Paglione06}. The same general trend is true of YbRh$_2$Si$_2$ in Fig.~\ref{fig:2}c. The low-field curves in YbRh$_2$Si$_2$ show a feature below 100~mK that is consistent with the presence of scattering associated with the antiferromagnetic ordering at $T_N$ = 80~mK, but as the field is increased the temperature dependence reverts to linear for $B > 300$~mT, approaching quadratic for the highest field ($B =1$~T) \cite{Pfau12}. 

Perhaps the most conspicuous difference in comparing heat transport in these materials is in the magnitude of $\delta(T)$.  This is shown versus the scaled temperature axis of $T/T_K$ in Fig.~\ref{fig:2}d, where $T_K$ is the respective characteristic energy scale (Kondo temperature) for each material.  The values for $T_K$ used are YbRh$_2$Si$_2$: $T_K \sim 25$ K \cite{Sichel03}, CeCoIn$_5$: $T_K \sim 35$ K \cite{Nakatsuji02} and $\beta$-YbAlB$_4$: $T_K \sim T_0 \sim 200$ K \cite{Matsumoto11}. At the lowest temperatures $T/T_K < 10^{-3}$, $\delta(T) \rightarrow 0$, however  $\delta(T)$ increases more slowly with temperature in $\beta$-YbAlB$_4$ indicating a comparatively smaller amount of inelastic scattering in this material.  This is not simply due to differences in impurities levels between the systems, as CeCoIn$_5$ has a much lower residual resistivity ($\rho_0 = 0.1~\mu\Omega~\text{cm}$ \cite{Paglione06}) while for YbRh$_2$Si$_2$ the samples have a higher scattering rate ($\rho_0 = 1.6~\mu\Omega~\text{cm}$ \cite{Pfau12}). 

       \begin{figure} \centering
               \resizebox{\columnwidth}{!}{
\includegraphics{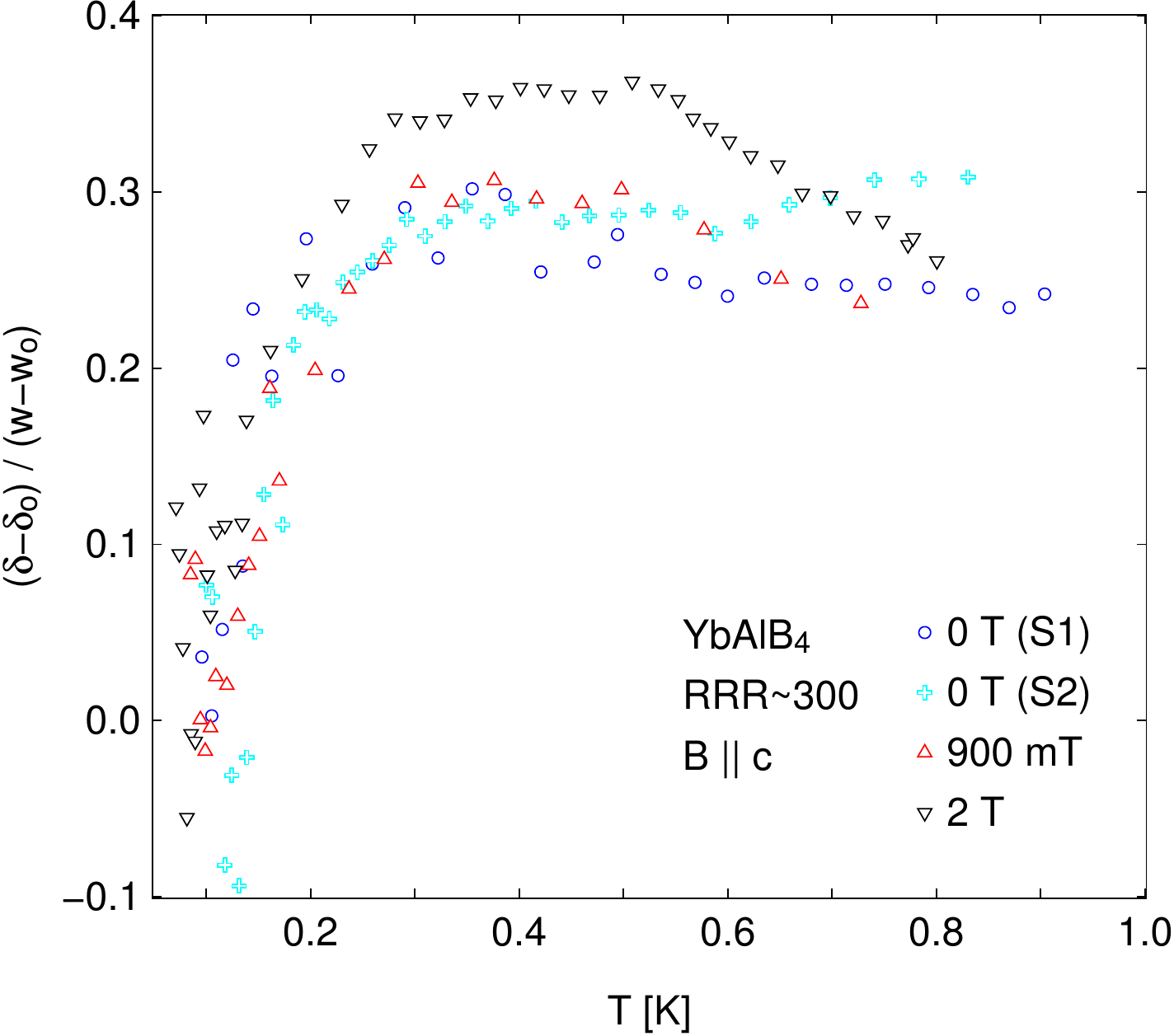}}
              \caption{\label{fig:3}The inelastic Lorenz ratio, $L_{in}$ = ($\delta-~\delta_0)~/~(w-w_0)$, in $\beta$-YbAlB$_4$, measured at several fields with $B\parallel c$.  The inelastic ratio, $L_{in}$, is a measure of the ratio of vertical (energy non-conserving) to horizontal (energy conserving) contributions to inelastic scattering processes. The data for two samples (S1 and S2) at zero field is shown.}             
      \end{figure}

Further insight may be gained by subtracting the elastic scattering and looking solely at the temperature dependence of the inelastic scattering channel. Defining the inelastic Lorenz ratio $L_{in}$ = ($\delta-\delta_0)/(w-w_0$), where $\delta_0$ and $w_0$ are the $T$ = 0 values of $\delta(T)$ and $w(T)$, allows access to the relative weighting of vertical to horizontal inelastic scattering processes, as $L_{in} \propto \frac{K_{ver}}{K_{hor}+K_{ver}}$. 
In a Fermi liquid, $L_{in}$ $\simeq$ 0.4-0.6, indicating the ratio of vertical to horizontal inelastic processes is close to unity and is roughly temperature independent \cite{Schriempf69,Kaiser71}. Fig.~\ref{fig:3}, however, reveals a remarkable feature of our data. A new energy scale, which is field independent, appears at a temperature of $T \sim$ 0.3 K below which the $inelastic$ Lorenz number $L_{in}$ falls rapidly \cite{note2}.  

Two conclusions can be drawn from the feature at $T \sim$ 0.3 K.  First, in conjunction with our observations of the magnitude of $\delta(T)$, the magnitude of $L_{in}$ above $T \sim 0.3$ K  indicates that both horizontal and vertical inelastic processes are small compared to that seen in similar quantum critical systems.  Second, the decrease in $L_{in}$ indicates a decrease in the ratio of vertical to horizontal inelastic processes as the temperature is lowered. Consequently, either the horizontal scattering processes are increasing, or the vertical processes are reduced, perhaps through a gapping of the fluctuation spectrum. This feature persists through the critical regime, and corresponds roughly to the temperature at which a $T^2$ Fermi liquid state is recovered in transport measurements \cite{Matsumoto15}. Low temperature inelastic neutron scattering studies would help establish what drives the collapse in $L_{in}$, and its significance to quantum criticality.

One possible explanation that accounts for these observations is the existence of two types of carriers in $\beta$-YbAlB$_4$, arising from the two bands that cross the Fermi level \cite{Tompsett10}. One is relatively fast and light and dominates transport, with only weak inelastic scattering arising mainly from a largely field independent scattering mechanism. Another carrier is slow and heavy and is strongly scattered by inelastic fluctuations, with a reasonable contribution to specific heat but a limited contribution to transport. There is some experimental support for this picture. The multiband Fermi surface revealed by experimental and theoretical studies \cite{OFarrell09,Nevidomskyy09} shows cyclotron masses ranging from $m^{\star}/m_0$ $\sim$ 3.6 - 13.1, for instance. Also, the entropy associated with the approach to the QCP in $\beta$-YbAlB$_4$ is quite small, with an upturn in $(C/T)T_0$ an order of magnitude smaller than in CeCu$_{5.9}$Au$_{0.1}$ or YbRh$_2$Si$_2$ \cite{Matsumoto11}, suggesting that perhaps only some electrons are renormalized in the approach to the critical point. Recent detailed Hall effect measurements \cite{O'Farrell12} are indeed well fit by a two-component model with one high mobility component dominating transport, at least at intermediate temperatures.

Based on our thermal transport observations, there are clearly three issues unique to $\beta$-YbAlB$_4$ that must be resolved with theoretical input. The first is the linear temperature dependence of $\delta(T)$ and its small magnitude in comparison to other field-tuned quantum critical systems, the second is the origin of the energy scale (~0.3 K) observed in the inelastic scattering channel and the third is the overall insensitivity of these transport phenomena to the applied magnetic field, perhaps arising from the low energy scales $T/T_K$ accessed in this system.

\begin{acknowledgments}

This research was supported by NSERC of Canada, the Royal Society, and the EPSRC.  This work was partially supported by Grants-in-Aid (No. 25707030) from the Japanese Society for the
Promotion of Science. E.O'F. also acknowledges support from the JSPS. We would like to thank Makariy Tanatar, Malte Grosche, Gil Lonzarich, Sven Friedemann and Mike Smith for useful discussions.
\end{acknowledgments}
\bibliographystyle{apsrev}
\bibliography{YbAlB4-kappa-final}

\begin{thebibliography}{38}
\bibitem[{not({\natexlab{a}})}]{note1}
\bibinfo{note}{Strictly speaking, the presence of superconductivity in the
  charge channel excludes any statement on the WF law in zero field in the
  $T=0$ limit. However, in a small magnetic field ($B$ = 50~mT), the
  superconducting transition in the resistivity is removed and we recover the
  WF law within our experimental resolution.}
\end{thebibliography}

\end{document}